\documentclass[12pt]{iopart}

\usepackage{graphicx}
\begin{document}

\title{Rheology of Hard Glassy Materials}

\author{A. Zaccone$^{1,}$$^2$ \& E. M. Terentjev$^2$}
\address{$^1$ Department of Physics ''A. Pontremoli", University of Milan, via Celoria 16, 20133 Milano, Italy}
\address{$^2$ Cavendish Laboratory, University of Cambridge, JJ Thomson Avenue, CB30HE Cambridge, U.K.}

\ead{az302@cam.ac.uk, emt1000@cam.ac.uk}
\vspace{10pt}
\begin{indented}
\item[]May 2020
\end{indented}

\begin{abstract}
\noindent Glassy solids may undergo a fluidization
(yielding) transition upon deformation whereby the material starts to flow plastically. It has
been a matter of debate whether this process is controlled by a specific time scale,
from among different competing relaxation/kinetic processes. Here, two constitutive models
of cage relaxation are examined within the microscopic model of nonaffine elasto-plasticity. One (widely used) constitutive model implies that the overall relaxation rate is dominated by the fastest between the structural ($\alpha$) relaxation rate and the shear-induced relaxation rate. A different model is formulated here which, instead, assumes that the slowest (global) relaxation process controls the overall relaxation. We show that the first model is not compatible with the existence of finite elastic shear modulus for quasistatic (low-frequency)
deformation, while the second  model is able to describe all key features of deformation of `hard' glassy solids, including the yielding transition, the nonaffine-to-affine plateau crossover, and the rate-stiffening of the modulus.
The proposed framework provides an operational way to distinguish between `soft' glasses and `hard' glasses based on the shear-rate dependence of the structural relaxation time. 
\end{abstract}

%
%
%
%
%

\section{Introduction}

Liquids behave like solids at sufficiently high rates of deformation, but at
very slow (quasistatic) deformation they flow with zero energy
cost~\cite{Frenkel,Trachenko}. Glassy solids exhibit a very similar
behaviour at intermediate to high deformation rates, but they possess a finite
 shear modulus when subject to quasistatic deformation. However,
when the amplitude of applied deformation is sufficiently large, glassy solids yield to
plastic deformation~\cite{Cates,Ballauff}.
The question about the kinetics of this yielding, or the elastic-plastic
transition, has a long history: it goes back to, at least, the work of
Eyring~\cite{Eyring}, who introduced the basic concepts still in use
today~\cite{Rottler,Barrat_review}.

The key concept in Eyring's theory and most of the subsequent treatments,
including  the Shear Transformation Zone (STZ) theory~\cite{Langer} or the 
Cooperative Shear Model (CSM)~\cite{Samwer1,Samwer2}, is that the
plastic flow sets in when the applied stress suppresses the barrier for
molecular
jumps out of the local energy well, such that the motion of molecules in the
direction of shear matches the globally applied shear rate. In such models,
there is a single relaxation time, 
set by the escape rate out of the energy well, and
often referred to as the $\alpha$-relaxation~\cite{Zaccone_review}.

Successive modifications of the Eyring model accounted for the distribution of relaxation times, following from the
distribution of energy wells~\cite{Bouchaud}. This approach led to the celebrated Soft Glassy
Rheology (SGR) theory~\cite{Sollich}. A different way of incorporating the
heterogeneity of the dynamical process is through the already mentioned STZ
model, which has proven useful in the modelling of real solids.

 In most of these models the relevant time scales for relaxation in the strained system are identifiable with \textit{local} energy barriers, although in Mode Coupling Theory the relaxation is more collective and cooperative~\cite{Cates,Ballauff}. Recently, it has been
emphasized that the dynamics of glasses may be controlled by local, rather than
global relaxation processes~\cite{Wyart}. Accordingly, these models cannot give closed-form constitutive relations depending on the overall (observable) structural relaxation time and the (externally imposed) deformation rate.

Here  we compare and contrast the `soft' and the `hard' glassy materials, where the soft glassy systems age, restructure, and therefore adjust their modulus on the experimental time scale. The `hard glass' instead is controlled by frozen-in configurations that are stable on a time scale much longer than any experiment: such amorphous solids appear with a well-defined plateau modulus at low frequencies.
We follow a different approach to the strain- and strain rate-dependent
deformation of glassy solids~\cite{ZacconePRB2014,Laurati2017}: our model is analytically tractable, and based on
the theory of nonaffine elasticity~\cite{Lemaitre,Zaccone2011,Zaccone2011b}.
The key role is played by the overall structural relaxation time.

 Within this elasto-plastic model, two different
relations for the structural relaxation time $\tau_{\alpha}$ are examined: one in which $\tau_{\alpha}$ is controlled 
by the slowest macroscopic process in the glass under dynamic shear, 
and the other, where it is controlled by the process with the highest rate, typically the local
shear-assisted bond-breaking time-scale. It turns out that only the former model can recover the hallmark of hard glassy solids: a non-zero shear modulus plateau at vanishing frequencies/rates. Other characteristics of glassy deformation including the yielding (elastic-plastic transition) and its temperature dependence, are also recovered.

The title of this paper is a deliberate counterpoint to the Soft Glassy Rheology theory~\cite{Sollich}, because we focus on the elasticity and yielding transition of true solids with the quasistatic shear modulus, whereas in SGR the long-time limit is that of a fluid flow.

\section{Nonaffine elastoplastic model}
\subsection{Free energy of deformation}
We start from a phenomenological model that describes the
mechanical response of glasses~\cite{ZacconePRB2014}.
The shear modulus for a generic amorphous
solid can be written in a form: $G=\frac{2}{5\pi}(\kappa\phi/R_{0})(z-z_{c})$. Here,
$\kappa$ is the bond spring constant, $\phi$ is the atomic/particle packing
fraction, $R_0$ is the mean distance between nearest
neighbours, and $z$ is the average number of mechanically-active 
neighbours. This mean contact number $z$ does not include nearest-neighbours
which are fluctuating fast in and out of contact.
As illustrated in many previous studies, the shear modulus vanishes at a
critical connectivity value $z_c$, due to nonaffine displacements which soften
the elastic
response~\cite{Zaccone2011,Zaccone2011b}. We recall that nonaffine
displacements are prominent in amorphous solids: due to the lack of local
symmetry in the particle environment, the forces that this particle receives
from its nearest neighbours in the affine position prescribed by the strain
tensor are unbalanced~\cite{Lemaitre}. This lack of mechanical
equilibrium in the affine position causes the additional displacement of the
particle
towards a true equilibrium position under the action of the unbalanced local force.
This additional displacement is accompanied by a
decrease in the free energy of deformation because the displacement implies a
mechanical work which is done by the solid to keep equilibrium.

This is true for both athermal solids (such as jammed packing of particles) and
also for thermal systems (e.g. polymer or metallic glasses).
If interparticle interactions are purely central-force, then
$z_c=6$~\cite{Thorpe1}, reflecting the celebrated Maxwell counting of constraints, whereas
for more complex interactions one has $z_c=2.4$ for covalent
networks~\cite{Thorpe2,Zaccone2013_cov}, and $z=4$ for a glass of linear
polymer chains~\cite{Zaccone2013,Lappala,Hoy} (where a mixture of covalent bonds and 
central-force Lennard-Jones type interactions is present).

The elastoplastic free energy $F_{el-pl}$ response to the imposed shear deformation $\gamma$ can be
written as: 
$F_{el-pl}(\gamma)=F_{A}(\gamma)-F_{NA}(\gamma)$
with two contributions
corresponding to the affine deformation (as in Born-Huang theory and its
extensions~\cite{Born}), and to the nonaffine deformation (the sum of negative
local, internal work contributions), respectively.
Using the generic shear modulus $G(z)$, this free energy becomes
\begin{equation}
F_{el-pl} = \frac{1}{2} \left( \frac{2 \kappa \phi}{5\pi R_0} \right) \big[z(\gamma)-z_{c} \big]\gamma^{2},
\label{free} 
\end{equation}
  {where the modulus $G(z)$ incorporates the microscopic parameters of spring constant $\kappa$, and the mean packing fraction $\phi$.} The source of elastic nonlinearity here is the change (reduction) of the mean contact number $z$ with increasing deformation in the affine part of
$F_{el-pl}$~\cite{ZacconePRB2014}.
It is important to point out that the nonaffine part of this deformation free energy remains quadratic in the strain amplitude $\gamma$, even beyond the yield point of the glass. This theoretical prediction~\cite{Lemaitre,Zaccone2013_cov} has also been confirmed experimentally in colloid glass by measuring the mean squared nonaffine displacement at different strains~\cite{Laurati2017}.

\begin{figure}
\centering
{\includegraphics[width=0.75\columnwidth]{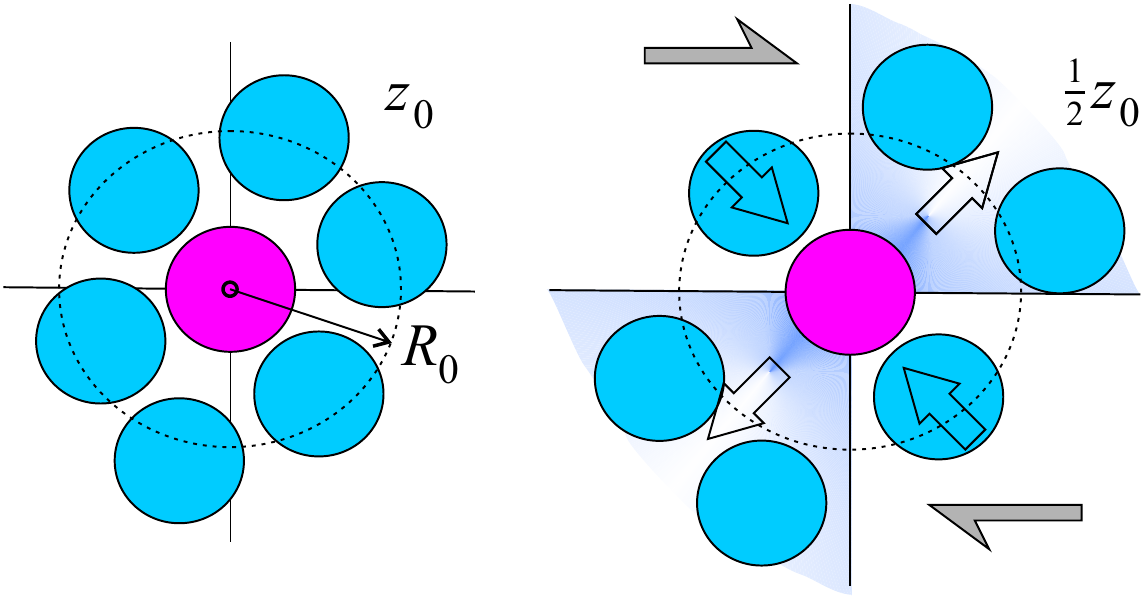}
\caption{(Color online) A scheme of local cage packing, when the two extension
sectors under applied shear reduce the number of mechanically-active contacts,
while the two compression sectors do not change their mean connectivity. }
\label{fig1sectors}}
\end{figure}

As discussed in studies by various authors, upon being sheared, the
glassy cage of nearest-neighbours gets deformed in such a way that neighbours
are lost in the two extensional sectors of the solid angle around a given
particle~\cite{Bouchaud,ZacconePRB2014}.
In the two compression sectors, the particles, instead, are pushed against
the test particle, but due to excluded volume there is practically no gain of
new mechanical contacts, see Fig. \ref{fig1sectors}. Hence, there is a net
decrease of the total $z$ due to
the shear deformation. In many amorphous solids, this is associated with dilatancy~\cite{Tighe},
although in some glasses the dilatancy could be too small to be measured as the decrease of $z(\gamma)$ can be
associated with redistribution of particles in free volume pockets.

\subsection{Evolution of connectivity with strain}
 The simplest physically-motivated expression that captures the relevant limits of undeformed glass $z\rightarrow z_{0}$, and the post-yielding fluidized state $z \rightarrow z_{0}/2$ (which would be equal to 6 in the Maxwellian packing of monodisperse spheres where the local shear modulus becomes zero~\cite{Zaccone2011}),
with an exponential crossover dependence on both the energy barrier (Arrhenius) $\beta \Delta \approx T_{g}/T \geq 1$~\cite{ZacconePRB2014}, and on the shearing time, is as follows:
\begin{equation}
z(\gamma)=\frac{{z}_{0}}{2}\left[1+e^{-(T_{g}/T)\gamma}e^{-t/\tau_{\alpha}}\right],
\label{first}
\end{equation}
 where the factor $e^{-t/\tau_{\alpha}}$ is expected from the general solution to the time-dependent diffusive (Smoluchowski) dynamics \cite{Dhont}. This equation compactly expresses the fact that the number of long-lived neighbors $z$ decays (eventually to about a half of its value, in the two compacted sectors out of the four in Fig. \ref{fig1sectors}) either upon externally driving the system to a very large deformation $\gamma$, or simply by waiting for a very long time $t$ much longer than the characteristic $\alpha$-relaxation time, $\tau_{\alpha}$. Note that lowering the temperature much below $T_g$ makes the transition happen at smaller $\gamma$, i.e. the material is more brittle.
 Please note that within this picture, conceptually consistent with MCT~\cite{Goetze,Reichman}, the cooperatively-enhanced local cage-level energy barrier sets the global energy barrier, hence $\beta \Delta \approx T_{g}/T$ used above.


The factor $z_{0}$ is the mean number of mechanical contacts at rest; its value is
$z_{0}\approx 12$ for dense glasses~\cite{Hansen}, as also confirmed
experimentally in colloidal glass~\cite{Laurati2017}.
In a glass of  linear polymer chains, $z_{0}\approx 8$ was found in
Brownian-dynamics simulations~\cite{Hoy}, and by examining the packing of
tetrahedral sublattices~\cite{diMarzio}. Finally, in a silica glass (an
archetype system of random covalent bonds), the mean packing number was
found to be $z_0 \approx$ 5-6~\cite{Soules,Tse}.  Here we want
to propose that, as the mean connectivity $z_0$ of a hard glassy material diminishes
on increasing shear amplitude, its value does not drop below the critical
number at marginal stability $z_c$: At this point (when $z=z_c$) the solid is fully fluidised,
and any further decrease in density is unjustified.
Accordingly, we take $z_0/2 \approx z_c$. This assumption is supported
by many observations: in the colloid packing
$z_c=6$~\cite{Thorpe2,Laurati2017}, in a linear polymer $z_c=4$~\cite{Hoy,Lappala} and
in a covalent-bonded network $z_c=2.4$~\cite{Thorpe2,Zaccone2013_cov}. 

\subsection{Stress-strain relation and strain-dependent modulus}
 Putting Eq. (\ref{first}) back into the free energy of deformation, Eq.(\ref{free}), replacing $t$ with $\gamma = \dot{\gamma} t$ for the case of deformation ramp, and differentiating, the
following stress-strain relation for the non-viscous part of the stress under a
constant-rate strain ramp is obtained:  
\begin{equation}
\sigma = \frac{1}{4} z_{0} K \gamma~
e^{-A \gamma } \left(2- A\gamma \right) ,   \label{stress}
\end{equation}
with the linear modulus $G_0= \frac{1}{2} z_{0} K$, where we defined $K=\frac{2 \kappa \phi}{5\pi R_0}$. Predictions of this model have been found in quantitative
agreement
with experimental data on metallic glass~\cite{Johnson}, colloidal glass under
shear~\cite{Laurati2018,Egelhaaf,Laurati2017} and data of 2D colloidal glass at
the air-water interface \cite{Langevin,Maestro}.

Finally, upon differentiating the stress in Eq. (\ref{stress}), we 
find the expression for the modulus
\begin{equation}
G =\frac{1}{2} G_0 e^{-A\gamma}  \left(2-4A \gamma +A^{2}\gamma^{2} \right)
\label{modulus}
\end{equation}
where the shorthand parameter  $A=(T_{g}/T) + (\dot{\gamma}\tau_{\alpha})^{-1}$ is coming from Eq. (\ref{first}), and $\tau_{\alpha}$ is the structural relaxation time, which will also depend on the applied shear-rate $\dot{\gamma}$.

In the next section, we will address different conceptual models which express the overall $\tau_{\alpha}$ in terms of the underlying kinetic processes. We will see that different ways of combining the underlying rate processes leading to different $\tau_{\alpha}(\dot{\gamma})$ expressions, result in two completely different scenarios which can distinguish between "soft" and "hard" glasses.

\section{Rate-dependent relaxation time}

Equation (\ref{stress})  still does not take into account how the shear rate affects 
the structural relaxation time $\tau_\alpha$, which appears in Eq. (\ref{first}).
When the rate of a physical process is determined by the interplay of
sub-processes, each with its own kinetics, two possibilities exist:
either the overall relaxation rate is
controlled by the fastest of the two rates (summation in series), or instead it is
the longest relaxation time that controls the overall process, making the rates add in parallel. 

In our case, we could have the in-series addition of two key rates:
\begin{equation}
\tau_\alpha^{-1}=\tau_0^{-1} + \frac{\dot{\gamma}}{\gamma_{c}} \ , \label{series}
\end{equation}
where $\tau_0$ is the static structural $\alpha$-relaxation time of the cage,
and $\gamma_{c}$ is a constant
parameter that sets the amount of strain needed to
break a cage (typically $\gamma_{c} \sim 0.1$ ~\cite{Johnson,Voigtmann}). This constitutive expression is
saying that the faster rate of cage breaking controls the overall relaxation rate.
This relation is used within an extended version of Mode-Coupling theory
for sheared liquids, which is able to describe shear-thinning of viscoelastic
liquids~\cite{Voigtmann}. Furthermore, this relation has been found in both experiments and simulations of supercooled liquids~\cite{Onuki} and polymer melts~\cite{Hoshino,Douglas}.

One should note that in the static limit
$\dot{\gamma}\rightarrow 0$ one recovers $\tau_\alpha \rightarrow \tau_0$, with
a finite cage relaxation time, as typical for liquids. 
Note that a similar relation to Eq. (4) with a power-law exponent $n$ acting on $\dot{\gamma}$ defines a broader class of Herschel-Bulkley models ~\cite{Herschel}.

A different choice would be to say that the longest process time dominates
the overall relaxation dynamics
 (in a solid this would be the structural relaxation). In this case, the
in-parallel summation gives
\begin{equation}
\frac{1}{1/\tau_\alpha} = \frac{1}{1/\tau_0} + \frac{1}{\dot{\gamma}} \ , \
\ \ \mathrm{i.e.} \ \
\tau_\alpha^{-1}=\frac{\dot{\gamma}}{1+\dot{\gamma}\tau_0} ,  \label{parallel}
\end{equation}
where $\tau_0$ is the equilibrium cage relaxation.  That is, we measure the total time of relaxation in two steps: $\tau_0+1/\dot{\gamma}$, as e.g. in the reaction-diffusion case, where the total time of the process is a sum of the two consecutive times. When $\dot{\gamma}$ is high, the deformation is \textit{affine} (not able to relax local internal forces by adjusting positions and reducing $z_0$ connectivity). An affine deformation also implies that the characteristic time of the structural rearrangement is much larger than the time-scale of the external driving, $\tau_0 \dot{\gamma} \gg 1$. In general, $\tau_0$ could be identified with the Maxwell-Frenkel relaxation time, approximately given by the hopping time of one atom to get out of the cage~\cite{Trachenko}. Since in a high-rate affine deformation the atoms never leave the cage (their motion is limited to high-rate motions within the cage), it is clear that the time-scale set by $\tau_0$ is the one which governs the structural relaxation.

In contrast, at low $\dot{\gamma}$ such that $\tau_0 \dot{\gamma} \ll 1$, the relative internal positions adjust non-affinely into much deeper minima under the action of the external drive, and the stronger barriers to relaxation result in the increase of the effective $\tau_\alpha$. 
However, for hard materials, in practice, the overall relaxation time is close to the equilibrium structural relaxation time $\tau_0$. For example, for silicate glasses, the structural relaxation time is found experimentally on the order of 19,000 years~\cite{Mauro} so that $1/\tau_0$ is probably much smaller than any experimentally accessible $\dot{\gamma}$.

\begin{figure}
\centering
{\includegraphics[width=0.75\columnwidth]{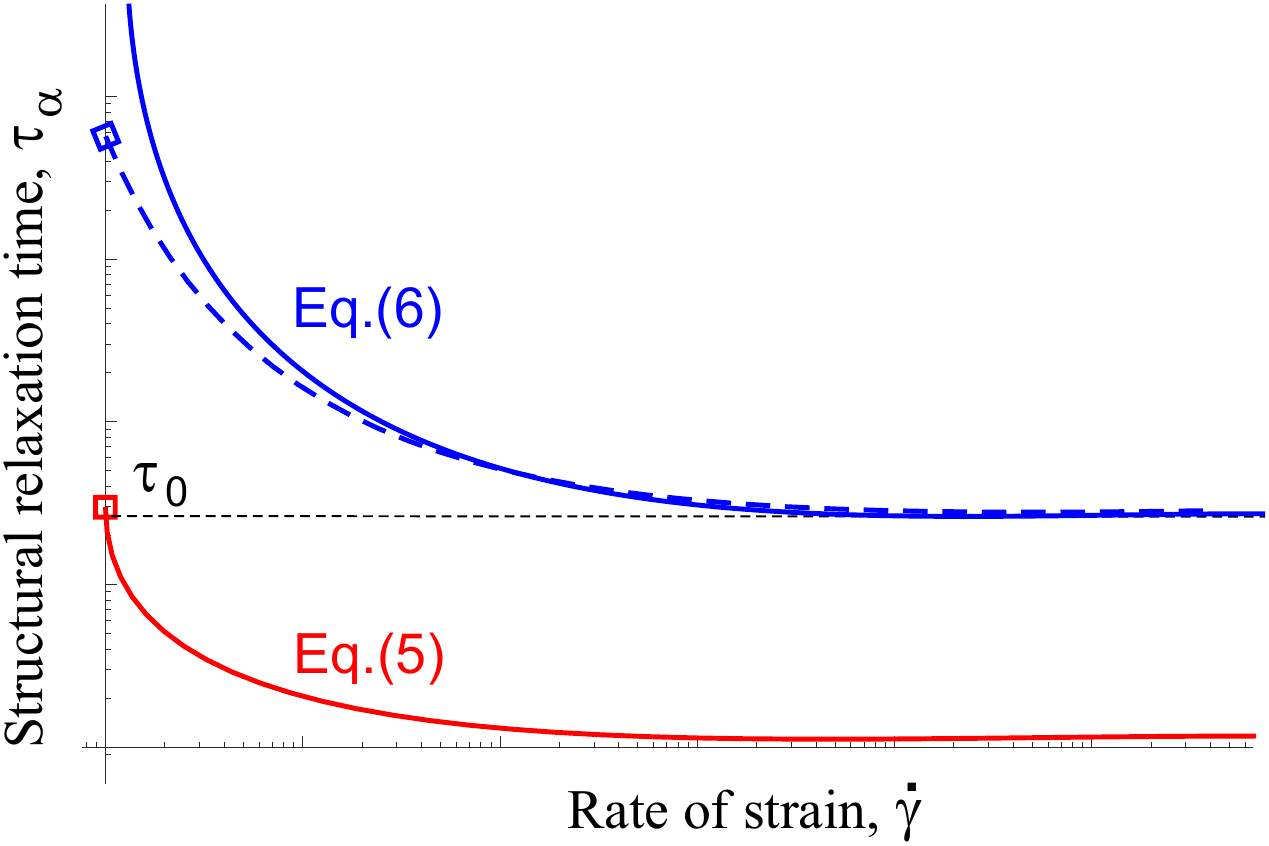}
\caption{(Color online) A scheme comparing the addition of relevant rates according to the in-series model (liquid-like, Eq.~5), and in-parallel model (solid-like, Eq.~6), with reference to the equations in the text. The finite relaxation time at $\dot{\gamma} \to 0$ is a feature of SGR model, i.e. the liquid-like quasistatic response, presented as dashed line to contrasti it with the prediction of the in-parallel addition model.}
\label{fig2compare}}
\end{figure}

The two models are schematically compared in the cartoon of Fig. \ref{fig2compare}.

\section{Results and comparison}
\begin{figure}
{\includegraphics[width=0.85\columnwidth]{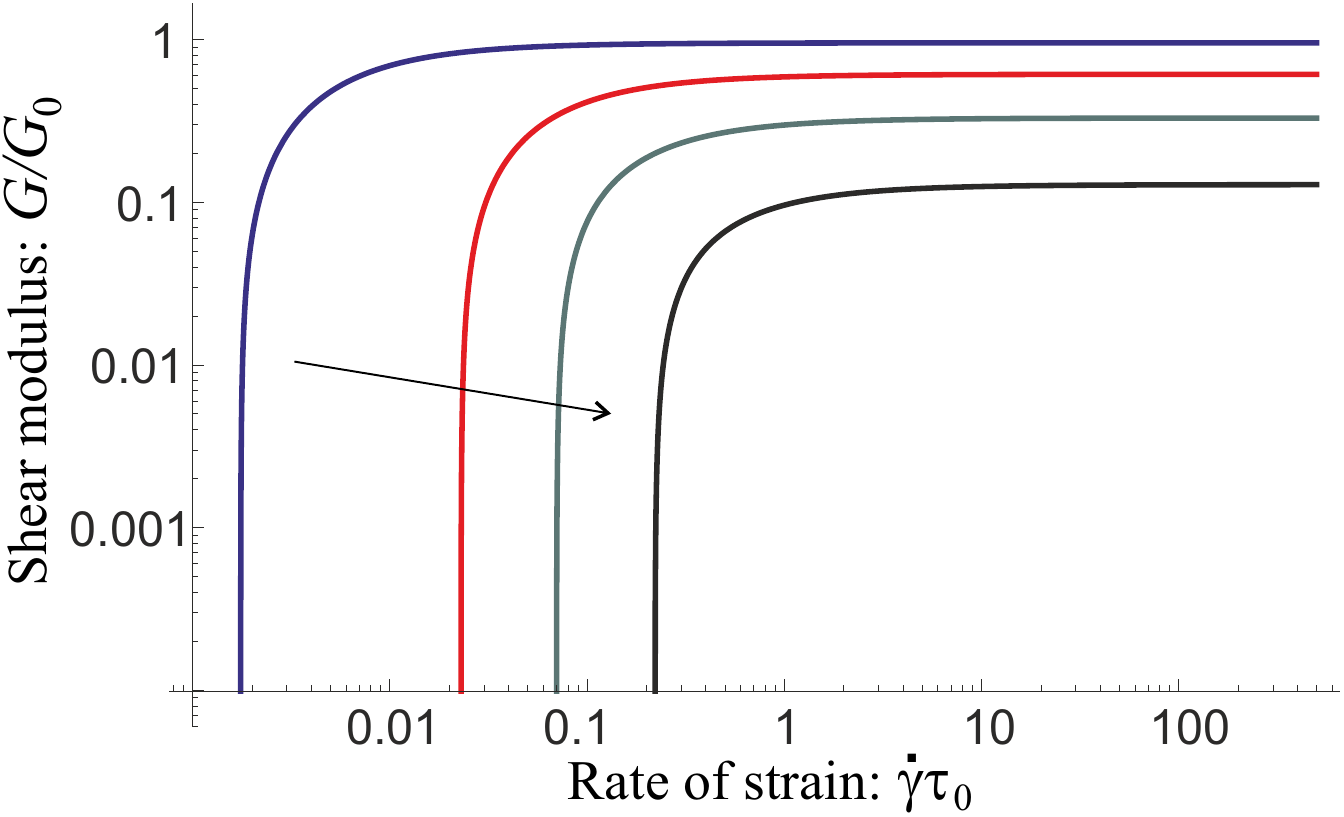}
\caption{(Color online) Model predictions from Eq. (\ref{modulus}) using the
constitutive
relation Eq. (\ref{series}). For the calculations the cage barrier has been
kept constant,
 $\beta\Delta=5$.
 From top to bottom the applied strain is:
$\gamma=0.001, \, 0.01, \, 0.02, \, 0.03$.  {The parameter $\gamma_c =0.1$ (strain at cage breaking) is kept constant for all curves.} }
\label{fig2wrong}}
\end{figure}

We will now compare predictions of the model with these two different constitutive
relations for the relaxation process. In Eq.(\ref{modulus}), we replace  $\tau_\alpha=\gamma_c \tau_0 /
(\gamma_c+\dot{\gamma})$ inside the parameter $A=(T_{g}/T) + (\dot{\gamma}\tau_{\alpha})^{-1}$, which implements the in-series rate addition in Eq.(\ref{series}). This modulus is plotted in Fig. \ref{fig2wrong} as a function of increasing shear rate, for several values of the final strain $\gamma$.
It is clear that the solid-like response for quasistatic
 deformation ($\dot{\gamma}\rightarrow 0$) is never attained, even at
the smallest strain. This shows that the constitutive model given by Eq.
(\ref{series})
cannot describe a solid in equilibrium, but at most a viscoelastic liquid. We should
also remark on the fact that this outcome is in contradiction with the underlying premise
of the theoretical model which assumes the existence of a free energy of
deformation quadratic in the strain, Eq. (\ref{free}).

\begin{figure}
\centering
{\includegraphics[width=0.85\columnwidth]{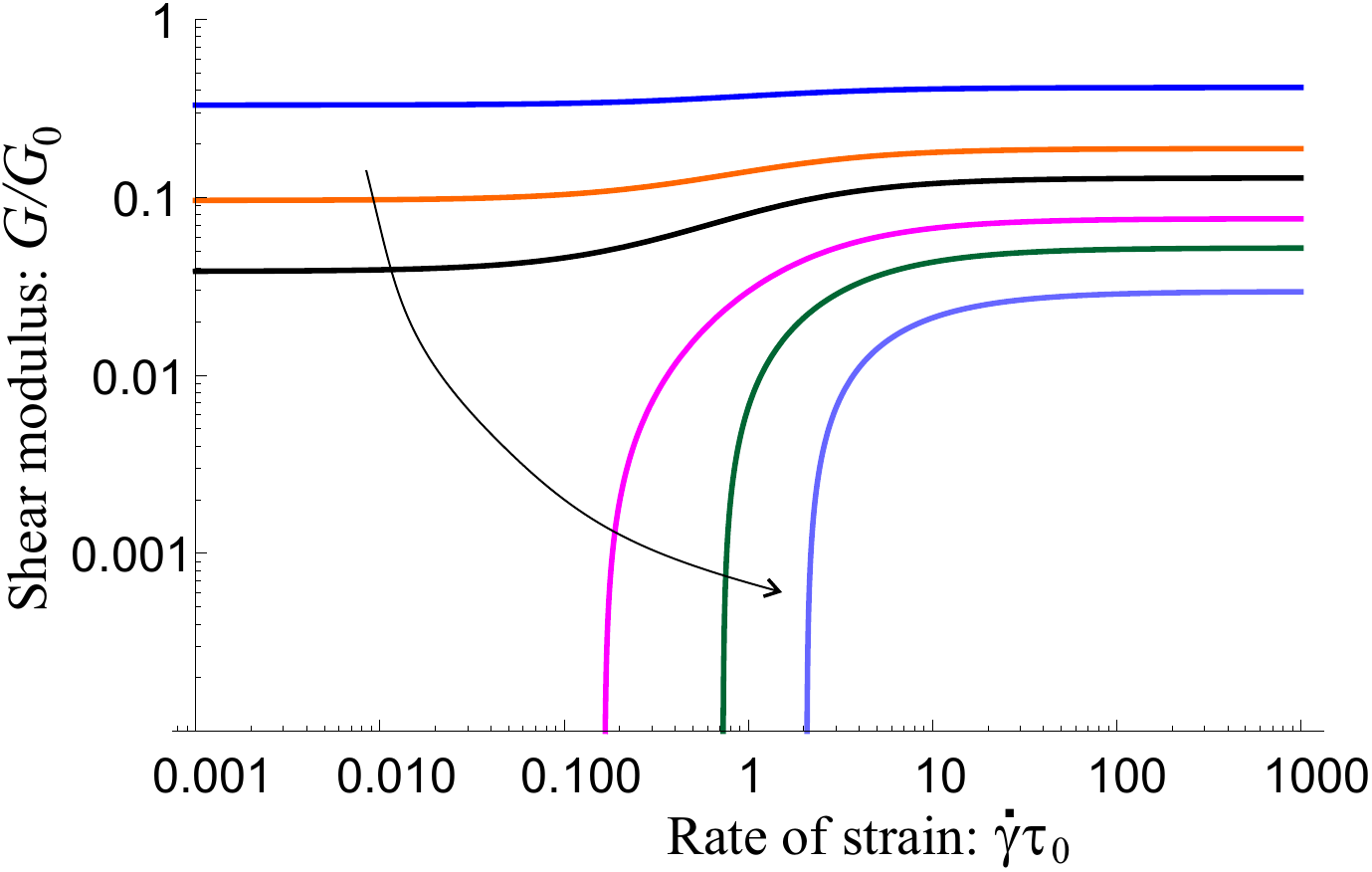}
\caption{(Color online) Model predictions from Eq. (\ref{modulus}) using the
constitutive
relation Eq. (\ref{parallel}). The cage barrier   $\beta\Delta=5$.
 From top to bottom the applied strain is:
$\gamma=0.05, \, 0.08, \, 0.09, \, 0.1, \, 0.105, \, 0.11$.  {The parameter $\gamma_c =0.1$ (strain at cage breaking) is kept constant for all curves.}  }
\label{fig3right}}
\end{figure}

In contrast, when we apply the in-parallel constitutive model for the rates
addition, $\tau_\alpha=(1+\dot{\gamma}\tau_0)/\dot{\gamma}$ in Eq. (\ref{parallel}), the plots in Fig.
\ref{fig3right} show the shear modulus having the correct qualitative
behaviour. At very low  strain, the modulus reaches the quasistatic
(equilibrium) plateau. However, above a critical yield
strain $\gamma^*$ the low-rate response is at zero modulus (plastic flow). This
elastic-plastic 
crossover is achieved because the structural relaxation diverges in the
constitutive model of Eq. (\ref{parallel}). The low-rate (equilibrium) plateau
is dominated by the nonaffine dynamics, after which a smooth crossover leads to a
higher plateau at high strain rates, which is instead affine.

\begin{figure}
\centering
\includegraphics{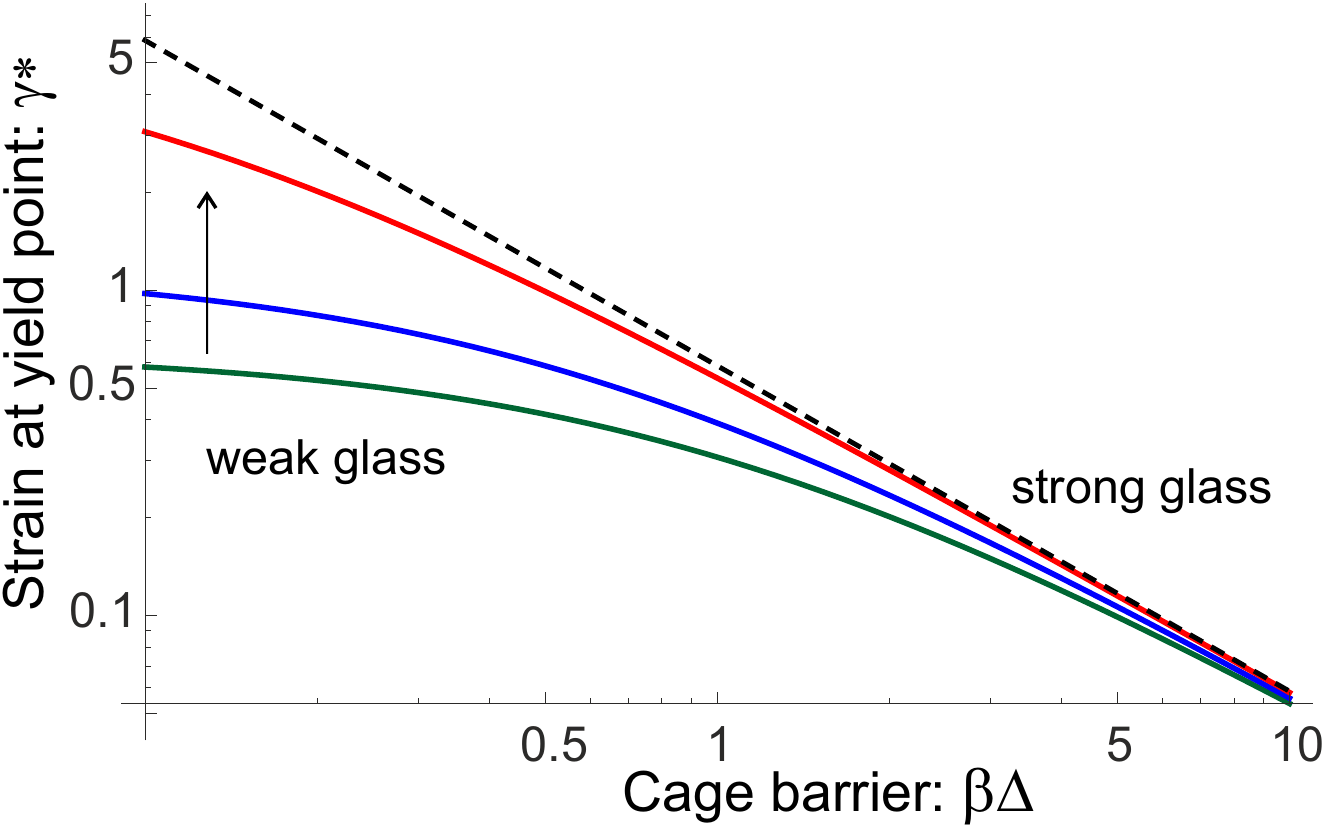}
\caption{(Color online) The dependence of the yield strain $\gamma^*$ on the
strength of cage barrier  $\beta\Delta = T_g/T$, for different rates of strain ramp.
The dashed line is showing the asymptote $\gamma^* = (2-\sqrt{2})/\beta \Delta$.
 From top to bottom the curves are for increasing shear rate:
$\dot{\gamma} \tau_0 = 0.1, \, 1$ and $10$. }
\label{fig4yield}
\end{figure}

Figure \ref{fig4yield} shows how the yield point $\gamma^*$  (the solution of
$\partial \sigma/\partial \gamma =0$) approaches the simple asymptotic value
$\gamma^* = (2-\sqrt{2})/ \beta\Delta$, independent of the strain rate when the cage barrier is high. That is,
strongly bonded glass reaches the limit of its elastic response already at a very small
strain -- while in a weak glass the yield point has a strong dependence on the
rate of applied shear, expressed as $\dot{\gamma} \tau_0$ in Fig. \ref{fig4yield}. 

\begin{figure}[t]
\centering
{\includegraphics[width=0.85\columnwidth]{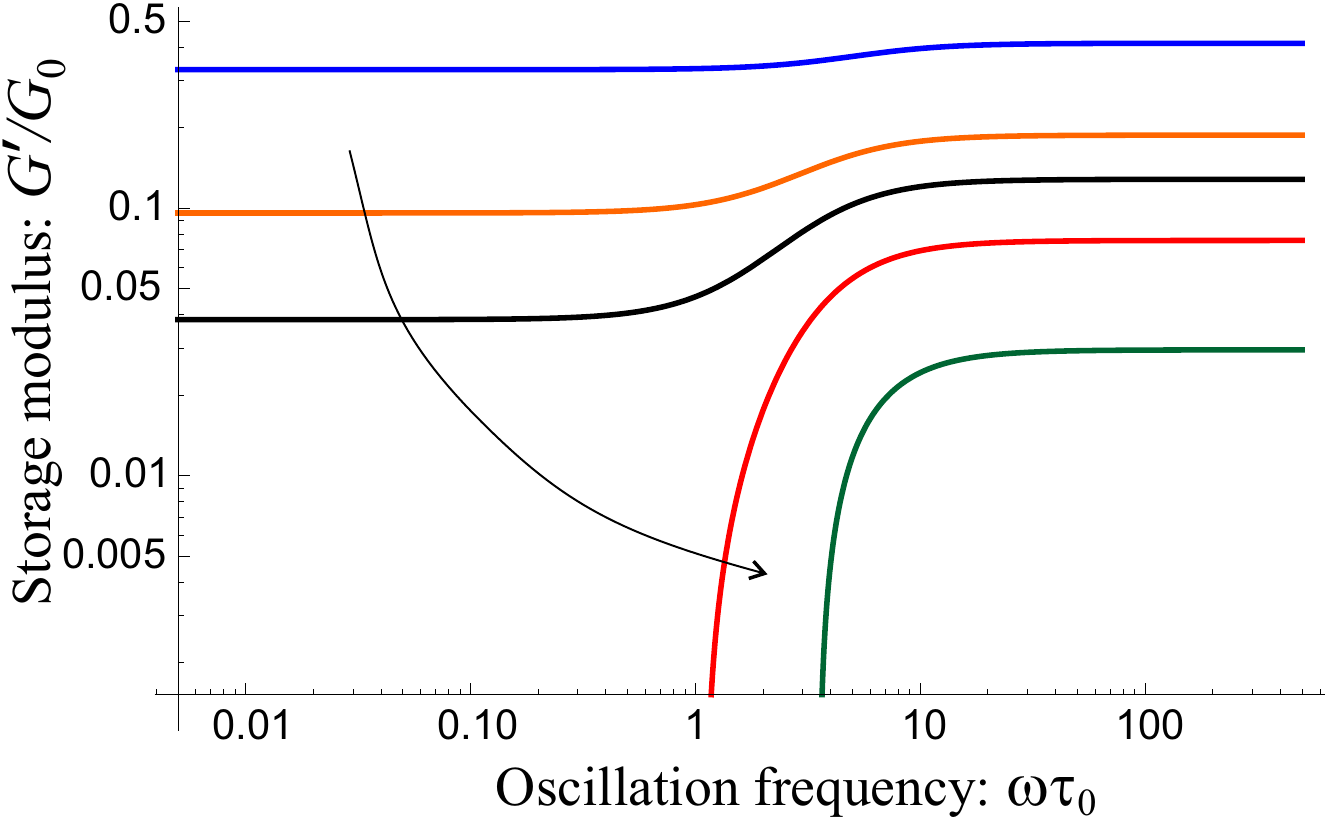}
\caption{(Color online) The storage modulus $G'(\omega)$, for  $\beta\Delta=5$
and for different amplitudes of oscillating strain: $\gamma_{0} = 0.05, \, 0.08, \,
0.09, \, 0.1$ and $0.11$, from top to bottom, again revealing the fluidisation
above a critical strain amplitude.  {The parameter $\gamma_c =0.1$ (strain at cage breaking) is kept constant for all curves.} }
\label{fig5real}}
\end{figure}

The same qualitative behaviour of $G$ is to be expected for the
complex modulus $G^*(\omega)$ in response to an oscillating deformation of
amplitude $\gamma$.
It has been shown~\cite{Richeton,Xiao} that the rate-dependent modulus can be converted into a
frequency-dependent complex modulus merely by replacing $\dot{\gamma}$ with the
relation
$\dot{\gamma}\approx4 i \omega\gamma_0$, where $\gamma$
and $\omega$ are the wave amplitude and the frequency of the signal in
oscillatory shear deformation.
The resulting storage modulus $G'(\omega)$ is in full agreement with experimental data on
a variety of materials~\cite{Zener,Weitz,Federico,Carter,Wilde,Duffrene,Schaller}, all
featuring a low-frequency equilibrium modulus plateau (with strong nonaffine
behaviour) which transitions smoothly to an upper (affine) plateau at higher
frequency, Fig. \ref{fig5real}.  {Upon reaching a sufficiently high strain amplitude, the atoms in the extensional sectors of the cage (see Fig. 1) have now left the cage and the critical condition $z\rightarrow z_0/2$ is reached at which fluidization occurs, leading to the disappearance of the low-frequency plateau in the post-yielding regime. Although there is not yet experimental confirmation of this trend as a function of strain amplitude, similar curves of fluidization have been reported for $T$-induced fluidization as a function of $T$~\cite{Barrat,Wilde,Schaller}.}

\begin{figure}
\centering
{\includegraphics[width=0.65\columnwidth]{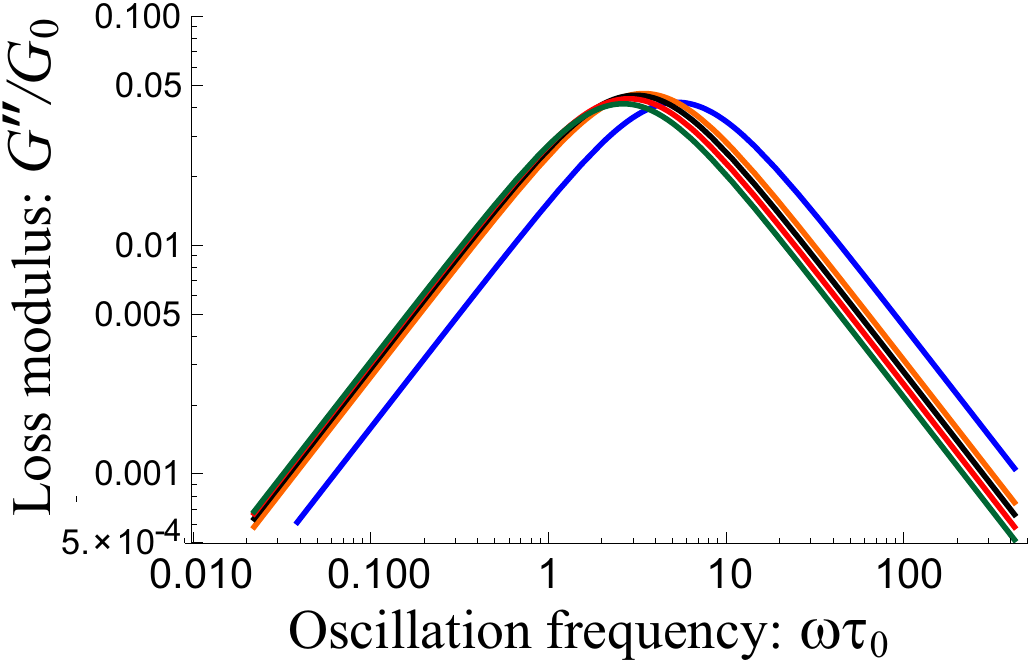}
\caption{(Color online) The loss modulus $G''(\omega)$, for $\Delta/k_BT=5$ and
strain amplitudes: $\gamma_{0} = 0.05, \, 0.08, \, 0.09, \, 0.1$ and $0.11$. The
relaxation peak at $\omega \tau_0 \sim 1$  changes very little during the
fluidisation transition.  {The parameter $\gamma_c =0.1$ (strain at cage breaking) is kept constant for all curves.}}
\label{fig6im}}
\end{figure}

The plots in Fig. \ref{fig6im} show the corresponding imaginary part of the complex modulus, $G''(\omega)$. It also shows an expected behaviour consistent with experimental trends on various hard glasses (metallic, silicates)~\cite{Wilde,Schaller,Duffrene}, with the characteristic peak in the loss modulus around the characteristic frequency $\omega \tau_0 \sim 1$, and a power-law decay on both sides of the peak. It is, however, unexpected to see almost no difference in the loss modulus for different strain amplitudes: in great contrast to the storage modulus that shows a very dramatic fluidisation effect. Again the small horizontal shift of the resonance peak in $G''$ is consistent with trends observed in hard glasses upon $T$-induced fluidization~\cite{Wilde,Schaller,Duffrene}. In retrospect, we have to accept that the loss mechanism in this theory arises from the cage re-arrangement sketched in Fig. \ref{fig1sectors}, which is the same microscopic process on either side of the elastic-plastic transition. 

\section{Conclusion}
Our main conclusion is that to describe a solid glass elasticity, one must adopt a physical view of microscopic cage dynamics expressed by the constitutive relation Eq.(\ref{parallel}). This implies that the overall cage relaxation time under an external dynamic strain is dominated by the internal parameter $\tau_0$ at shear rates  $\dot{\gamma} \tau_0 \geq 1$. In contrast, in quasi-static equilibrium, the glass behaves as a perfect solid with a well defined reference state, and the finite equilibrium shear modulus. The model reproduces the elastic-plastic transition, with the yield strain $\gamma^*$ in Fig. \ref{fig4yield} that distinguishes between the strong and weak glasses. In turn, the parameter which discriminates between strong and weak glass is the bonding energy barrier $\Delta$ for a nearest-neighbour to be removed from the cage. In the future, the spatial variation of energy barriers and of relaxation times can be added into the model to address more complicated strain histories and the specificity of various material chemistries.

\ack
Th. Voigtmann is gratefully acknwoledged for the useful discussion and input.  This work was supported by the US Army ARO Cooperative Agreement W911NF-19-2-0055 (AZ)  and the EPSRC via Theory of Condensed Matter Critical Mass Grant EP/J017639 (EMT).
\\

\end{document}